\documentclass{emulateapj}
\def\vel{km s$^{-1}$}

\slugcomment{}

\shorttitle{\sc IRC+10216 narrow lines}
\shortauthors{\sc Patel et al.}

\begin{document}


\title{Submillimeter narrow emission lines from\\ the inner envelope of
IRC$+$10216}


\author{Nimesh A.\ Patel\altaffilmark{1}, Ken H.\ Young\altaffilmark{1},
Sandra Br\"unken\altaffilmark{1},
Robert W.\ Wilson\altaffilmark{1}, \\ Patrick Thaddeus\altaffilmark{1}, Karl
M.\ Menten\altaffilmark{2}, Mark Reid\altaffilmark{1}, Michael C.
McCarthy\altaffilmark{1},\\
Dinh-V- Trung\altaffilmark{3,4}, Carl A. Gottlieb\altaffilmark{1}, Abigail
Hedden\altaffilmark{1} }

\altaffiltext{1}{Harvard-Smithsonian Center for Astrophysics,  60 Garden
Street, Cambridge, MA; npatel@cfa.harvard.edu}
%, kyoung@cfa.harvard.edu, rwilson@cfa.harvard.edu,
%sbruenken@cfa.harvard.edu, pthaddeus@cfa.harvard.edu,
%mreid@cfa.harvard.edu, mmcarthy@cfa.harvard.edu, cgottleib@cfa.harvard.edu,
%ahedden@cfa.harvard.edu}
%KMM
\altaffiltext{2}{Max-Planck-Institut f{\" u}r Radio Astronomie, Auf dem H{\"
u}gel 69, D-53121, Bonn, Germany}
\altaffiltext{3}{Academia Sinica Institute for Astronomy and Astrophysics,
Taipei, Taiwan}
\altaffiltext{4}{On leave from Institute of Physics, Vietnamese Academy of
Science and Technology, 10 Daotan, Badinh, Hanoi, Vietnam}

\begin{abstract}
A spectral-line survey of IRC+10216 in the 345 GHz band has been
undertaken with the Submillimeter Array.  Although not yet completed,
it has already yielded a fairly large sample of narrow molecular
emission lines with line-widths indicating expansion velocities of
$\sim$4 \vel, less than 3 times the well-known value of the 
terminal expansion velocity (14.5 \vel) of the outer envelope.
Five of these narrow lines have now been  identified as rotational
transitions in vibrationally excited states of previously detected
molecules: the v=1, J=17--16 and J=19--18 lines of Si$^{34}$S and
$^{29}$SiS and the v=2,  J=7--6 line of CS.  Maps of these lines
show that the emission is confined to a region within $\sim$60 AU
of the star, indicating that the narrow-line emission is probing
the region of dust-formation where the stellar wind is still being
accelerated.

\end{abstract}


\section{Introduction}
Massive circumstellar envelopes of Asymptotic Giant Branch (AGB)
stars are believed to be a major contributor of molecules and grains
to the interstellar medium. IRC+10216 (CW Leo), at a distance of
150 pc,  is the archetypal AGB carbon star with a high mass-loss
rate ($> 10^{-5}$ M$_{\odot}$ yr$^{-1}$)  (Young et al. 1993; Crosas
\& Menten 1997). Owing to its proximity, this star is an ideal
target for detailed studies of physical and chemical processes in
AGB circumstellar envelopes  (e.g. Olofsson et al. 1982). Nearly 60 molecular
species have been discovered in  its circumstellar shell from
single-dish spectral-line surveys (Kawaguchi et al. 1995; Avery et
al. 1992; Groesbeck et al. 1994; Cernicharo et al. 2000; He et al.
2008).

Interferometric mapping allows us to check predictions of abundances
of various molecular species as a function of radius in the 
circumstellar envelope. ``Parent'' molecules, such as CO, C$_2$H$_2$, CS,
HCN, \& SiS,
are formed in the stellar atmosphere in thermo-chemical equilibrium
(Tsuji 1964, 1973). Once they get levitated
to a distance from the star at which the density is too low for
chemical reactions, their abundances ``freeze out'' at the values prevailing
in that
region (McCabe et al. 1979). At around that distance ($\sim20$ AU), the
temperature  has dropped below the dust condensation value ($\sim1200$ K)
and
dust grains start forming (Monnier et al. 2000).
Radiation pressure accelerates the grains and, by friction, the molecules too,
in an outflow which, for IRC+10216, reaches
a terminal velocity of 14.5~km~s$^{-1}$.
In the outer parts of the expanding envelope, a rich carbon-dominated photo
chemistry driven by the ambient ultraviolet field
produces species such as  CN and C$_{4}$H
whose mostly optically thin emission can be
observed as ring-like distributions with radii, at a few times
$10^{16}$--$10^{17}$ cm
depending on the chemical reactions at work (see the reviews of Glassgold
1996;
Ziurys 2006).

At submillimeter wavelengths, we can observe lines
requiring elevated excitation conditions, allowing us to probe the
physical
conditions in  the inner circumstellar envelope (radius $\lesssim10^{16}$
cm) where densities and temperatures are relatively high (temperatures
$\sim$100--1000 K, column densities $\sim 10^{22}$--$10^{24}$ cm$^{-2}$)
(Keady \& Ridgway 1993).

We have begun a spectral-line survey of IRC+10216 with the Submillimeter
Array\footnote{The Submillimeter Array is a joint project between
the Smithsonian Astrophysical Observatory and the Academia Sinica
Institute of Astronomy and Astrophysics, and is funded by the
Smithsonian Institution and the Academia Sinica.} (SMA, Ho et al.
2004).  Previous 345 GHz single-dish line surveys include those done with the
JCMT (Avery et al. 1992) and CSO (Groesbeck et al. 1994) in the
frequency ranges of 339.6--364.6 GHz and 330.2--358.1 GHz, respectively.
Our SMA line survey will eventually cover the frequency range of
300--355 GHz with higher sensitivity and spatial resolution than previous 
surveys; about 40\% of the survey has so far been completed. 
The frequency range 300--330 GHz  will be observed for the first
time. Here we present several new results, including the
discovery of many lines having narrow widths, and detections of
vibrationally excited rotational transitions in SiS, $^{29}$SiS,
Si$^{34}$S and CS.  A full account of the observed lines will
be presented on completion of the survey.

\section{Observations and Data reduction}\label{obs}

We observed IRC$+$10216 with the SMA in 2007 February in the
subcompact configuration with baselines from 9.5 m to 69.1 m in the
frequency range of 300 to 355 GHz.  To follow up on some of the
detected narrow lines with higher angular and spectral resolution,
we repeated the observations at 337.5 GHz with the SMA in the
extended configuration on 2008 February 19. The baseline lengths
in this configuration range from 44.2 m to 225.9 m. The synthesized
beam sizes were $3''\times2''$ and $0.''8\times0.''6$ in the
subcompact and extended array observations, respectively.  Table 1
summarizes the observational parameters on the five tracks of
observations which are relevant for the data presented here. The
duration of each track was from 7 to 9 hours. The phase center was
$\alpha(2000)=09^{h}47^{m}57.38^{s}, \delta(2000)=+13^{\circ}16'43.''70$
for all observations. All the tracks in subcompact configuration
were carried out in mosaiced mode, with 5 pointings with offsets
in RA and DEC: $(0'',0'')$ and $(\pm12'',\pm12'')$. The extended
configuration observations were carried out with single pointing
toward IRC+10216.  Titan and the quasars 0851+202 and 1055+018 were
observed  every 20 minutes for gain calibration.  The spectral
band-pass was calibrated using observations of Mars and Jupiter.
Absolute flux calibration was determined from observations of  Titan
and  Ganymede.

The visibility data were calibrated using the {\it Miriad} package
(Sault, Teuben \& Wright 1995). The mosaiced images are deconvolved
using the Miriad task {\it mossdi};  the resulting synthesized beams
are summarized in Table 1.  Maps of continuum emission show the
peak to have a position offset of
($\Delta\alpha,\Delta\delta)\approx(0.''7,0.''2)$ from the phase
center position quoted above.  The absolute position measurements
in the continuum emission are estimated to be accurate to $\sim
0.''1$.  Taking into account the proper motion of IRC$+$10216 of
($\dot{\alpha},\dot{\delta})\approx(26,4)$ mas yr$^{-1}$  determined
by Menten et al. (2006) our position is, within the mutual uncertainties
consistent with that determined by those authors.  The continuum
emission was unresolved at the highest angular resolution of $\sim
0.''8$.  The integrated continuum flux density was 0.84 Jy at 301.1
GHz and 1.17 Jy at 337.5 GHz, with an uncertainty of about 15\% in
the absolute flux calibration.
 All the spectra shown below were produced by
integrating the continuum-subtracted line intensity in a $2''\times2''$
rectangle centered on continuum peak (by means of the Miriad task
{\it imspec}).

\section{Expansion velocities}\label{histogramVexp}

Within a 3$''$ beam a total of 92 lines were detected in the first
phase of the SMA line survey of IRC+10216 at the central position.
Truncated parabolic line profiles were fitted using the CLASS package
(using the {\it shell} model for the line-profile) with one of the
fitted parameters being  $V_{exp}$, the expansion velocity of the
circumstellar shell.  From this sample, 25 lines have $V_{exp} \le$
7 \vel and of these, 12  are as yet unidentified. Several are
tentatively identified to be lines of salts such as KCN, NaCl and
NaCN.  Some of the lines may result from known molecules in
vibrationally excited states, such as the v$_{3}$=1 $15_{7,9}-14_{7,8}$,
$15_{7,8}-14_{7,7}$ doublet of SiCC at 345727.3 MHz.  Figure 2 shows
a sample spectrum toward IRC+10216 observed on 2008 February 9 over
a 2 GHz wide band centered at 337.5 GHz.  A comparison with the
line survey  of Groesbeck et al. (1994) (see their Figure 1) shows
that only the C$^{34}$S J=7--6 line at 337.396 GHz was detected in
their observations. All of the new narrow lines were missed in this
previous survey due to poorer sensitivity.  Here we present results
on the lines which are securely identified as rotational transitions
in vibrationally excited states of molecules that are well known
to be abundant in IRC+10216's envelope.

In previous single-dish line surveys of IRC+10216 (Cernicharo et
al. 2000; Kawaguchi et al. 1995; Avery et al. 1992; Groesbeck et
al. 1994) $V_{exp}$ is not tabulated, but it can be seen from
published spectra to be $\sim$14.5 \vel\ for all lines. From the
latest published line survey by He et al. (2008) (see their Table
9.), we can plot a distribution of $V_{exp}$ which is shown in
Figure 1 as empty bars with bold outlines. This histogram peaks at
14 \vel.  The distribution of $V_{exp}$ from our line survey is
shown in Figure 1 as grey bars.  Our line survey shows a peak in
the same bin of 14 \vel  but reveals a significant number of narrow
lines with velocities around $\sim$4 \vel.  Both histograms show a
continuous distribution of expansion velocities between these two
peaks at 4 and 14 \vel.

\begin{center}
\begin{figure}
\includegraphics[angle=-90,width=3in]{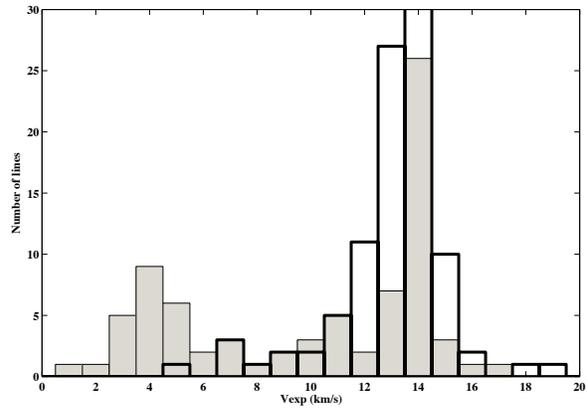}
\caption{Distribution of expansion velocities derived from lines detected
in the circumstellar envelope of IRC+10216. The uncertainty in $V_{exp}$  is
$\sim$0.2 \vel ($1\sigma$). White bars with bold outlines are from the
recent line survey of He et al.  (2008) which is  representative of all
single-dish line surveys toward IRC+10216 at cm to submm wavelengths. The
bin at 14 \vel  is shown truncated here; it actually consists of  170
lines. Grey bars represent SMA observations, showing a new population
of narrow lines which peaks at $\sim$4 \vel. }

\end{figure}
\end{center}

Lines with V$_{exp}\le 10$ \vel\ from IRC+10216 have been reported
by Highberger et al. (2000) and were assigned as vibrationally
excited SiS and CS lines.  He et al. (2008)  found four lines with
$V_{exp}$=7--10.2 \vel\ (see their Table 14) -- all from vibrationally
excited SiS.  Narrow maser lines in IRC+10216 from SiS and HCN were
reported by Fonfr\'ia-Exp\'osito et al. (2006) and Schilke \& Menten
(2003), respectively. Savik-Ford et al. (2004) detected OH -- which
has a narrow width (5.8 \vel), but this line appears at $-$37 \vel,
blue-shifted with respect to the systemic velocity of the star of
$-26$ \vel.  All the narrow lines we have observed are centered at
the systemic velocity of about -26 \vel to within $\sim$2 \vel.
Infrared molecular line profiles in the 10 $\mu$m band were observed
and analyzed by Keady \& Ridgway (1993).  They proposed (see their
Figure 3) an expansion velocity as a function of radius with: (1)
V$_{exp}=3\sim4$ km s$^{-1}$ from 1$\sim8$ R$_{*}$, (2) rapid
acceleration from 4 to 11 km s$^{-1}$ from 8 to 10 R$_{*}$ and (3)
slower acceleration from 11 \vel  to the terminal velocity of 14
km s$^{-1}$ from 10 to 20 R$_{*}$.  The histogram of expansion
velocities (Figure 1) peaks at $\sim$4 km s$^{-1}$ and $\sim$14 km
s$^{-1}$,  consistent with this velocity structure.

\begin{center}
\begin{figure}
\includegraphics[angle=-90,width=3.5in]{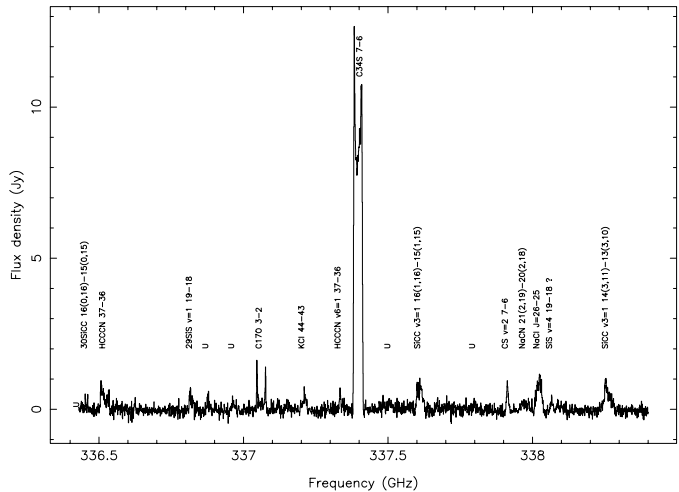}
\caption{A sample spectrum toward IRC+10216 showing several examples of narrow lines. Over this frequency range, the only line detected in a previous line survey (Groesbeck et al. 1994) was the C$^{34}$S J=7--6 line at 337.396 GHz. This line has a V$_{exp}$ =13.9 \vel. All the other lines are new detections and several of them are narrow (V$_{exp}<7$ \vel).}
\end{figure}
\end{center}

\section{Vibrationally excited SiS, $^{29}$SiS and
Si$^{34}$S}\label{vibexlines}

SiS is an abundant molecule in IRC+10216, one of the parent molecules
found close to the star (Glassgold 1996), and an important source
of Si for the formation of silicon carbides such as SiC, SiC$_{2}$
and SiC$_{4}$ (e.g., McCarthy et al. 2003).  Figure 3 (top) shows
the spectrum of SiS v=1 J=19-18 emission at 343100.98 MHz (rest
frequency), observed on 2007 February 12.  This line was previously
detected (but not mapped) in the single-dish line-survey of Groesbeck
et al.  (1994), where it is  significantly weaker than here because
of poorer sensitivity (we have longer integration time as well as
larger collecting area with the SMA).  Figure 3 (bottom) shows a
map of the integrated intensity emission,  over the velocity interval
of -40 to -10 \vel.  The total integrated flux density is 69.1 Jy
km s$^{-1}$.  The deconvolved source size is 1.1"x0.5" with P.A.
of $-27.6^{\circ}$. Assuming a size of $0.''5\times 0.''5$, we
estimate a lower limit to the brightness temperature of 200 K.

The emission appears to have an extended weak feature towards the
southeast.  High angular resolution near-IR adaptive-optics images
of IRC+10216 show evidence of azimuthally asymmetric structures in
the inner circumstellar envelope, over angular scales of $\sim2''$
(Menut et al. 2007). The 3$''$ angular resolution of our observations
is insufficient to allow a detailed comparison between the submillimeter
line emission maps and near-IR images.

The expansion velocity of the SiS v=1 J=19--18 line  is 10.6 \vel.
This is an example of a line of intermediate expansion velocity
(see Figure 1), in the rapidly accelerating zone of the envelope
where presumably dust has already formed.

Figure 4 shows spectra of Si$^{34}$S and $^{29}$SiS  v=1,  J=17--16
and 19--18 lines,  detected toward IRC+10216 for the first time.
This emission is unresolved and the source-size upper limits are
shown in Table 2. Assuming the size of the emitting region is
0.$''$2, we estimate  a  brightness temperature to be 140$\sim$200
K, which should be considered as lower limits.  We estimate a column
density of 7$\times10^{17}$ cm$^{-2}$ and abundance of 7$\times10^{-7}$
for $^{29}$SiS for an assumed excitation temperature of 550 K.

\begin{figure}
\includegraphics[angle=-90,width=3in]{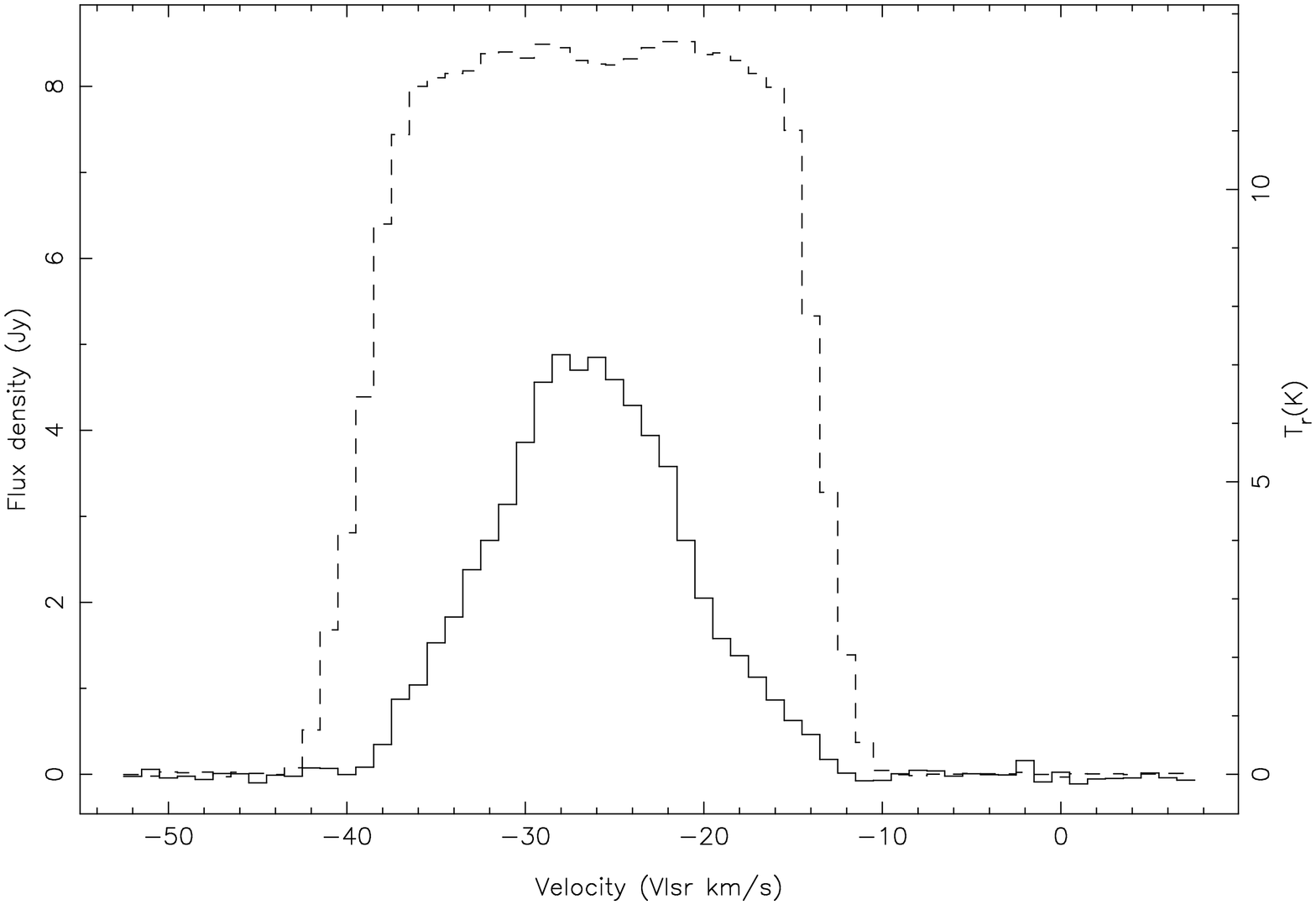}

\includegraphics[angle=-90,width=3.in]{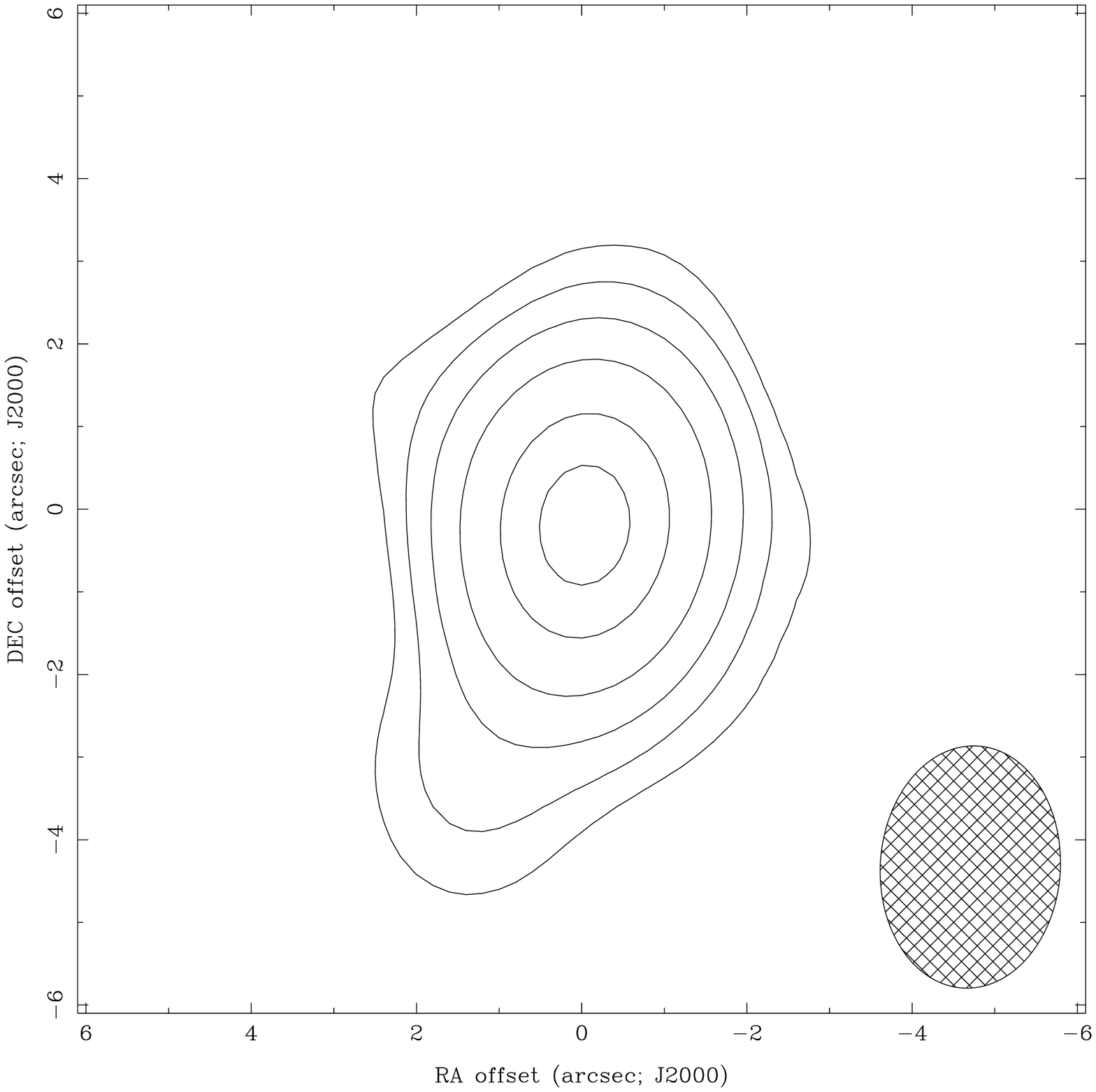}

\caption{Upper: Spectrum of SiS v=1 J=19-18 line obtained from a
2$''$ square centered on the star, with V$_{exp}$=10.6 \vel. For
comparison, the SiS, v=0 J=19--18 line is shown by the dashed line,
(scaled down by a factor of 10).  The v=0 line has an expansion
velocity of 13.5 \vel. Note that the intensity values for this line
are scaled down by a factor of 10 in this figure.   Lower: Integrated
intensity emission from the SiS v=1 J=19-18 line.  The contours
levels are -5, 5, 10, 20, 40, 80, 120 $\times$ 0.45 Jy/beam km/s.
The coordinate offsets are with respect to
$\alpha(2000)=09^{h}47^{m}57.43^{s},
\delta(2000)=+13^{\circ}16'43.''98$.The synthesized beam is shown
in the lower right corner. See also Table 2.}

\end{figure}

From the observed integrated intensities of SiS, $^{29}$SiS and Si$^{34}$S
v=1,  J=19--18 emission, we derive the isotopic abundance ratios
$[^{28}\rm Si/^{29}Si]=15.1\pm0.7$ and $[^{32}\rm S/^{34}S]=19.6\pm1.3$
(uncertainties are $1\sigma$). For comparison,   previously published
values from single-dish observations (Kahane et al. 1988; He et al.
2008) are $20.2\pm2$ for $[^{32}\rm S/^{34}S]$(in good agreement
with our estimate here) and $18.7\pm1$ for $[\rm Si/^{29}\rm Si]$ (marginally
larger than our value).  A possible cause for disagreement might be
non-negligible optical depths.

\begin{figure}
\includegraphics[width=3.in]{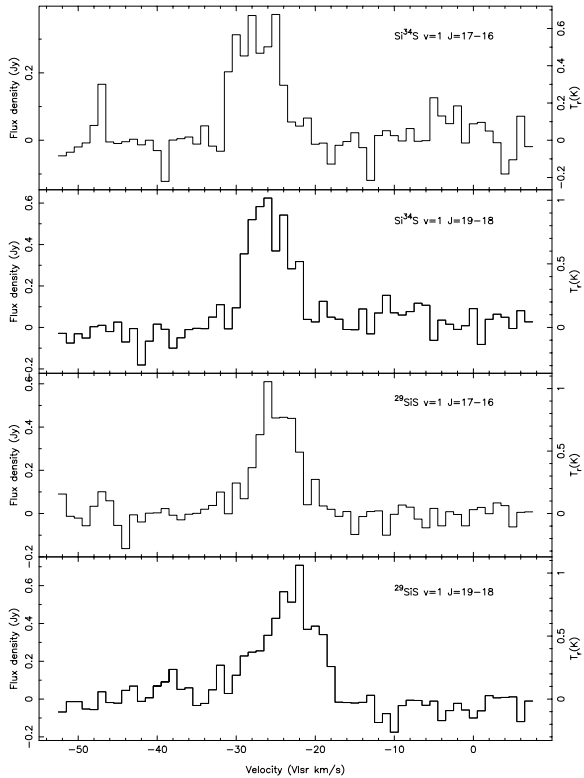}
\caption{Spectra of vibrationally excited rotational lines
in isotopes of SiS obtained from the imaged data cube. The flux density was
calculated over the area enclosed in a 2 $''$ square 
centered on the star.  See also Table 2.}
\end{figure}

\section{Vibrationally excited CS}

Vibrationally excited CS radio emission in IRC+10216  was first
detected by Turner (1987) in the v=1,  J=2--1 and 5--4 transitions.
These lines were re-observed along with new detections of the J=3--2,
6--5 and 7--6 transitions by Highberger et al. (2000). The J=7-6
emission reported here is the first detection in the v=2 state
(Figure 5). The triangular line-profile is indicative of spatially
unresolved emission from accelerating gas (Bujarrabal et al. 1986).
In subcompact configuration observations with angular resolution
of $\sim3''$, the emission appears highly concentrated and unresolved.
There is no emission detected at the angular radius of $\sim12''$
(where several other species have shown a peak in abundance in
previous interferometric maps). The observations of this line were
repeated in the extended configuration of the SMA at a beam size
of 0.$''$8, confirming that the emission is unresolved. The deconvolved
source size is $<0.''2$. The lower limit for brightness temperature
is 237 K. Assuming an excitation temperature of 550 K, we estimate
a column density of $7\times 10^{17}$ cm$^{-2}$ and a lower limit
for CS abundance with respect to H$_{2}$ of $9.3\times10^{-6}$
within a radius of $\sim 4.5\times10^{14}$ cm ($\sim 7 R_{*}$).

\begin{figure}
\includegraphics[angle=-90,width=3.in]{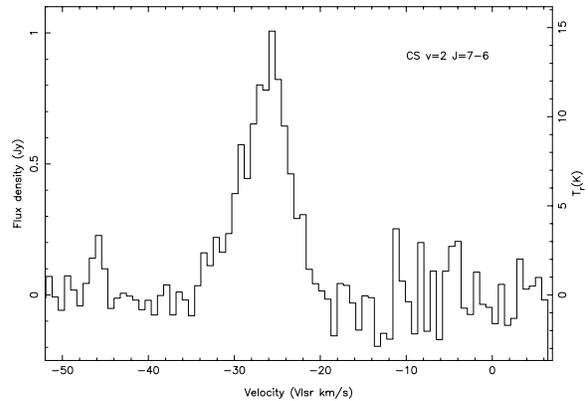}
\caption{Spectrum of CS v=2 J=7-6 emission at 337912.19 GHz. See Figure 4
caption and also Table 2.}
\end{figure}

Previous interferometric observations of the v=1, J=5--4 line show
the emission to be within a radius of $0.''35$ ($\sim$10 R$_{*}$
or $\sim 7\times10^{14}$ cm) (Lucas and Guelin 1999). Young et al.
(2004) derived a lower limit on the CS abundance to be $3.4\times10^{-9}$
within $\sim 34 R_{*}$. From single-dish observations of CS v=1,
J=3--2, 6--5 and 7--6 lines, Highberger et al. (2000) estimate an
abundance of 3--7$\times10^{-5}$ relative to H$_{2}$.  From comparison
with the CS radial abundance predicted by chemical models, we
conclude, as did Young et al. (2004), that the model of Millar et
al.  (2001) predicts too low a value ($\sim 10^{-11}$) at the radius
of $\sim 10^{16}$ cm.  In this model, assuming CS to be a parent
molecule (see Figure 1, right-panel, in Millar et al.  2001), the
initial abundance is still lower by about an order of magnitude
relative  to that derived from submillimeter observations. Moreover,
the drop in CS abundance with radius (owing to the production of
other sulphur-bearing molecules), is too small, and not consistent
with the compact distribution of CS seen in the SMA observations.
There is better agreement in a more recent study  of non-equilibrium
chemistry of the inner wind, which takes into account shocks induced
by stellar pulsation (Cherchneff 2006), although these models seem
more relevant for S stars (C/O$\approx 1$).

The CS v=2,  J=7--6 transition requires extreme excitation conditions
since it corresponds to an energy E$_{u}$/k=3707 K.  As noted by
Highberger et al. (2001), even for the v=1 line, collisional
excitation with H$_{2}$ would require very high gas densities
($\sim1-5\times10^{14}$ cm$^{-3}$. This line emission is most
plausibly excited by 8 $\mu$m stellar thermal radiation.

\section{Conclusions}\label{conclusions}

Preliminary results from the SMA line survey of IRC+10216 have
yielded a population of narrow lines with expansion velocities of
$\sim$ 4 \vel.  About half of these can be assigned to vibrationally
excited rotational transitions of abundant species such as CS, SiS
and their isotopomers. The emission is found to occur in a very
compact region smaller than $0.''2$ around the star.  This is thought
to be the region where dust is forming in the envelope and  in which
the material has just begun accelerating and has yet to attain the
terminal velocity of $\sim$14 \vel. The CS v=2,  J=7-6 line is most
likely radiatively excited,  because collisional excitation would
require an unrealistically  high gas density and abundance of CS.

\acknowledgments
It is a pleasure to thank Ray Blundell for his help and support
present on the SMA IRC+10216 line-survey.  We thank Mark Gurwell,
Thushara Pillai and Jun-Hui Zhao for helpful discussions on SMA
data reduction.  This research has benefitted from the Cologne
Molecular Spectroscopy Database (M\"uller et al. 2001; M\"uller et
al. 2005) (http://www.astro.uni-koeln.de/site/vorhersagen/), and
the ALMA group's spectral line catalog website:\\
http://www.splatalogue.net (Remijan et al. 2007) and the CASSIS
(Centre d'Analyse Scientifique de Spectres Infrarouges et
Submillim\'etriques) software (http://cassis.cesr.fr).

%\clearpage

\begin{deluxetable}{lccclcc}
%\rotate
\tablecolumns{7}
\tablewidth{0pt}
\tablecaption{Summary of observations\label{table1}}
\tablehead{
Date&SMA&Tuning&Synthesized&$\tau_{225 GHz}$&T$_{sys}$\\
      &configuration&&beam &  &(SSB, K)}
\startdata
2007 February 7&subcompact&299.1 GHz (LSB)&$3.''2\times2.''4$, P.A.=
-4$^{\circ}$&0.08&  180--270\\
2007 Feburary 8&subcompact&301.1 GHz (LSB)&$3.''3\times2.''4$,
P.A.=-6$^{\circ}$ &0.09& 160--230\\
2007 Feburary 9&subcompact&337.5 GHz (LSB)&$3.''0\times2.''4$,
P.A.=-8$^{\circ}$ &0.08& 180--350\\
2007 Feburary 12&subcompact&334.4 GHz (LSB)&$3.''0\times2.''2$,
P.A.=-3$^{\circ}$ &0.05& 130--260\\
2008 Feburary 19&extended&337.5 GHz (LSB)&$0.''8\times0.''6$,
P.A.=-88$^{\circ}$ &0.03& 100--250\\
\enddata
\end{deluxetable}
\normalsize

%\clearpage

\begin{deluxetable}{ccccccccc}
%\rotate
\tablecolumns{9}
\tablewidth{0pt}
\tablecaption{Summary of narrow lines\label{table2}}
\tablehead{
Species&Transition&Rest frequency\tablenotemark{1}&Catalog
freq\tablenotemark{2}.&V$_{exp}$&Peak&Integrated
&Deconvolved&T$_{B}$\tablenotemark{4}\\
            &                &          &          &  &flux
density\tablenotemark{3}&flux density&  size&\\
            &                & (MHz)             &  (MHz)         &  (km
s$^{-1}$)&(Jy)& (Jy km s$^{-1}$) & &(K)  }
\startdata
CS & v=2 J=7--6 & 337913.246 $\pm$ 0.18 & 337912.189 & 5.1$\pm$0.1 &0.78&
7.0 & $<0.''2$&$>$237.0\\
Si$^{34}$S & v=1 J=17--16 & 298630.441 $\pm$ 0.69 & 298629.989 & 4.5$\pm$0.4
&0.36& 2.6 &    $<1.''9\times 0.''2$&$>$138.7\\
%, P.A.=27.$^{\circ}$4\\
Si$^{34}$S& v=1 J=19--18 & 333733.581 $\pm$ 0.32 & 333731.998& 5.7$\pm$0.1
&0.49& 4.5 & $<2.''3\times 0.''5$&$>$183.3\\
%, P.A.=20.$^{\circ}$4\\
$^{29}$SiS& v=1 J=17--16 & 301390.406 $\pm$ 0.25 & 301388.939 & 4.8$\pm$0.2
& 0.52&3.8 & $<1.''9\times0.''8$&$>$198.3\\
%, P.A.=-18.3$^{\circ}$\\
$^{29}$SiS& v=1 J=19--18 & 336819.842 $\pm$ 0.32 & 336814.954 & 7.5$\pm$0.1
& 0.50&4.6 & $<0.''9\times0.''6$&$>$152.9\\
%, P.A.=-6.4$^{\circ}$\\
\enddata
\tablenotetext{1}{Assuming a systemic velocity of $-26.2$ \vel.}
\tablenotetext{2}{From the Cologne Database of Molecular Spectroscopy.}
\tablenotetext{3}{Typical uncertainty is 0.1 Jy.}
\tablenotetext{4}{\vspace{0cm}Lower limit in brightness temperature assuming
source size of $0.''2\times0.''2$}
\end{deluxetable}
\normalsize

\end{document}